# Scaling behaviors of RESET voltages and currents in unipolar resistance switching

S. B. Lee,[1] S. C. Chae,[1] S. H. Chang,[1] J. S. Lee,[2] S. Seo,[3] B. Kahng,[2] and T. W. Noh[1,a)]

[1]*ReCOE & FPRD, Department of Physics and Astronomy, Seoul National University, Seoul 151-747, Korea*

[2]*Department of Physics and Astronomy, Seoul National University, Seoul 151-747, Korea*

[3]*Samsung Advanced Institute of Technology, Suwon 440-600, Korea*

[a)] Author to whom correspondence should be addressed. Electronic mail: twnoh@snu.ac.kr

Unipolar switching phenomena have attracted a great deal of recent attention, but the wide distributions of switching voltages still pose major obstacles for scientific advancement and practical applications. Using NiO capacitors, we investigated the distributions of the RESET voltage and current. We found that they scaled with the resistance value $R_o$ in the low resistance state, and that the scaling exponents varied at $R_o \approx 30\ \Omega$. We explain these intriguing scaling behaviors and their crossovers by analogy with percolation theory. We show that the connectivity of conducting filaments plays a crucial role in the RESET process.

Although reversible resistance switching (RS) behavior induced by electric stimulus has been known since the 1960s,[1] interest in these intriguing physical phenomena has been renewed due to the potential applications for nonvolatile resistance random access memory (RRAM). Of particular interest is unipolar RS, in which switching occurs due to two applied voltages of the same polarity. Many consider unipolar RRAM a good candidate for multi-stacked, high density, and nonvolatile memory.[2] Unfortunately, unipolar RS is usually accompanied by wide distributions of the switching voltages, which make it difficult to fabricate reliable RRAM devices.[3] Despite their importance, we have little understanding of the switching voltage distributions.

Unipolar RS is widely accepted to occur because of the formation and rupture of conducting filaments under an electric field.[1] The switching voltage distributions may be closely related to the connection and disconnection of the conducting filaments. To describe collective changes in the conducting filaments' connectivity, we recently developed a new type of percolation model: the random circuit breaker (RCB) network model.[4] This model was able to successfully explain the reversible dynamic processes observed in the unipolar RS. The resulting percolating cluster was highly directional, unlike the nearly isotropic infinite cluster of classical percolation theory. However, our recent experimental work using third harmonic generation demonstrated that the connectivity of the conducting filaments in the unipolar RS could be quite similar to that in classical percolating systems.[5]

Here, we investigate how the percolating conducting filaments could become ruptured in one of the resistance switchings, called the RESET. In the unipolar RS, we show that the switching voltage and current scale with the resistance in the low resistance state. Depending on the resistance, there are two scaling regimes. We discuss the crossover between the two regimes in terms of the internal cluster structure of percolating filaments.

We grew a polycrystalline NiO thin film on Pt/Ti/$SiO_2$/Si substrates using dc magnetron reactive sputtering. Applying photolithographic techniques, we deposited 80-nm-thick Pt top electrodes with an area of 30 × 30 $\mu m^2$. The detailed fabrication methods for the Pt/NiO/Pt capacitors are described elsewhere.[3] Using a semiconductor parameter analyzer (Agilent 4155C; Agilent Technologies, Santa Clara, CA), we measured current–voltage (*I*–*V*) curves. To obtain reliable statistics, we obtained 100 *I*–*V* curves from seven NiO capacitors.

Figure 1 shows the 100 *I*–*V* curves of our NiO capacitors, which were typical unipolar RS. The NiO capacitors were highly insulating in pristine states. When we applied a voltage of approximately 5 V (not shown here), *I* increased suddenly, causing a complete dielectric breakdown. To prevent such permanent damage, we kept the current in the NiO capacitor below a compliance level $I_{\mathrm{comp}}$. Immediately after this Forming process, the capacitor entered a low resistance state (LRS). When $V$ was increased above $V_R$, it changed from the LRS to a high resistance state (HRS). We called this the RESET process. As we increased $V$ again in the HRS and reached $V_{\mathrm{S}}$, the film changed back into the LRS at a much higher voltage. We called this the SET process. For both Forming and SET processes, we used an $I_{\mathrm{comp}}$ of 1 mA.

The wide distributions of $V_R$ and $V_{\mathrm{S}}$ in Fig. 1 are very important. Here, we concentrate on the distribution of $V_R$ during the RESET process. In addition, we also examine the current value $I_R$ during the RESET process. As shown in the inset, we determined the resistance value $R_o$ of the LRS by fitting a linear line to the $I$–$V$ curve in the small $V$ region.

At first sight, the fluctuations in $V_R$ (Fig. 1) appear so large that it seems impossible to find any simple relation for $V_R$. However, when we plotted $\log_{10}(V_R)$ *vs.* $\log_{10}(R_o)$, we found that the $V_R$ data converged to two simple power law dependences. As shown in Fig. 2(a), $V_R \propto R_o^{-0.7 \pm 0.2}$ for $R_o < R_{\mathrm{co}}$, and $V_R \propto R_o^{+0.2 \pm 0.1}$ for $R_o > R_{\mathrm{co}}$, where $R_{\mathrm{co}} \approx 30$ $\Omega$. Similar scaling behaviors were also found for $I_R$, as shown in Fig. 2(b): $I_R \propto R_o^{-\alpha}$. The values of exponent $\alpha$ were $1.8 \pm 0.2$ for $R_o < R_{\mathrm{co}}$ and $0.7 \pm 0.1$ for $R_o > R_{\mathrm{co}}$.

To obtain a physical picture of each scaling regime, we chose eight points, marked the numerics in Fig. 2(a), and plotted the corresponding *I–V* curves. As displayed in Fig. 2(c), the *I–V* curves in the $R_o < R_{co}$ regime show a systematic decrease in $V_R$ (and $I_R$) with an increase in $R_o$. Similar behavior was reported in $TiO_2$ thin films[6] of 20–30 Ω resistance. However, as shown in Fig. 2(d), in the $R_o > R_{co}$ regime, a systematic increase in $V_R$ (and decrease in $I_R$) with the increase in $R_o$ occurred. Similar behaviors have been observed in Ti-doped NiO thin films[7] of $10^2$–$10^4$ Ω, and $Al_2O_3$ thin films[6] of approximately 100 Ω. Note that we were able to observe both results in our NiO capacitors.

How does the $\log_{10}(V_R)$ and $\log_{10}(I_R)$ data scale with $\log_{10}(R_o)$ in the RESET process? We must pay attention to the electrical breakdown process of semicontinuous metal films, which have been described by classical percolation theories.[8–10] As with our RESET process, the films were found to experience electrical breakdown at a threshold current $I_c$, when a hot spot reached the melting temperature of the metallic grains. Yagil, Deutscher, and Bergman found that $I_c \propto R_o^{-\alpha}$ with $\alpha = 1.75 \pm 0.4$ for $R_o < 2$ kΩ and $0.85 \pm 0.2$ for $R_o > 2$ kΩ.[9] These scaling behaviors in the two regimes and the values of $\alpha$ were very similar to our observations [Fig. 2(b)].

To obtain further insights, we measured the third harmonic generation response as a function of $R_o$. Details of the measurements have been reported elsewhere.[5,11] The third harmonic signal $V_{3f}$ probes the Joule heating in the percolating cluster in the LRS.[5,9–11] Figure 3 shows scaling behaviors of the third harmonic coefficient $B_{3f}$ ($\equiv V_{3f}/I^3$), with an applied ac voltage with a peak amplitude value of 0.1 V and a frequency of 10 Hz. As with the $V_R$ and $I_R$ data, $\log_{10}(B_{3f})$ in the LRS scaled with $\log_{10}(R_o)$. It also had two scaling regimes: $B_{3f} \propto R_o^{w+2}$ with $w = 4.7 \pm 0.3$ for $R_o < R_{co}$ and $w = 1.1 \pm 0.2$ for $R_o > R_{co}$.

For an inhomogeneous medium, the current distribution should be highly nonuniform. Moreover, $R_o$ measures the second moments of the current distribution in the network and $B_{3f}$ probes the fourth moments. Thus, $B_{3f}$ can be written as[5,10,11]

$$\frac{B_{3f}}{R_o^2} \propto \frac{\sum i_b^4}{\left(\sum i_b^2\right)^2} \propto R_o^w, \qquad (1)$$

where $i_b$ is the current flowing through each conducting filament $b$. Note that $1/f$ noise measurements also probe the fourth moments, so its scaling behavior can be also written in a form similar to (1). The experimental values of $w$ in the literature vary.[12] For two-dimensional semicontinuous Au films,[9] reports have indicated that $w \approx 1.65$ for $R_o < 2$ kΩ and $w \approx 0.1$ for $R_o > 2$ kΩ. For sand blasted films,[12] $w$ is reported to vary between 3.4 and 6.

Why are there two scaling regimes? According to classical percolation theory,[8,10] the percolating cluster can be viewed as a backbone composed of two groups. In one group, the bonds are in 'blobs,' which are multiply connected, but in the other group, the bonds are in 'links,' which are singly connected. If any bond in a link is cut, the backbone is split into two parts. The correlation length $\xi$ represents the average distance between the nodes, as shown in Fig. 4(a). The finite size $L$ of the sample can play an important role in determining the effective connectivity between two electrodes. As schematically shown in Fig. 4(b) [and 4(c)], if $\xi < L$ ($\xi > L$), all (some) parts of the infinite metallic cluster should be multiply (singly) connected. The connectivity change will cause changes in current distributions and the $w$ value.[8,10,13]

The connectivity changes between Figs. 4(b) and (c) can explain why the $I$–$V$ curves in Figs. 2(c) and (d) behave differently depending on the scaling regime. When $R_o < R_{\mathrm{co}}$ (i.e., $\xi < L$), all of the conducting filaments are multiply connected. When the number of the conducting channel decreases, the $R_o$ value increases. Because the current then flows along fewer channels, the Joule heating in each conducting channel increases significantly,[14,15] resulting in a decrease in $I_R$ and $V_R$. Our earlier simulations using the RCB network model confirmed this behavior.[4] When $R_o > R_{\mathrm{co}}$ (i.e., $\xi > L$), some parts of the conducting channel become singly connected. In this channel, all of the current should merge. Then, the increase in $R_o$ should be related to the length increase of the singly connected channel. Thus, $I_R$ should be nearly constant, and $V_R$ should increase. However, due to the Joule heating, $I_R$ could decrease with the $R_o$ value. Our recent simulation using the RCB network model with temperature dissipation effects also supports this argument. Details will be published elsewhere. Therefore, in our NiO capacitors, the scaling regimes of $R_o < R_{\mathrm{co}}$ and $R_o > R_{\mathrm{co}}$ should correspond to the regimes of $\xi < L$ and $\xi > L$, respectively.

In summary, we investigated the wide distributions of the RESET voltages and currents in NiO films with unipolar resistance switchings. We found that they scaled with the resistance value in the low resistance states. In addition, we found crossover behavior in the scaling. This intriguing behavior could be explained in terms of the connectivity of the percolating cluster.

This work was supported by the Creative Research Initiatives (Functionally Integrated Oxide Heterostructure) of the Ministry of Science and Technology (MOST) and the Korean Science and Engineering Foundation (KOSEF). B. K. and J. S. L were supported by the KOSEF grant funded by the MOST (No.R17-2007-073-01001-0). S. B. L. acknowledges support from a Seoul Science Scholarship.

**Figure captions**

FIG. 1. (Color online) One hundred current–voltage ($I$–$V$) curves from seven NiO capacitors, which show unipolar resistance switching. The red and blue lines correspond to the RESET and SET processes, respectively. To prevent permanent dielectric breakdown, we used the compliance current $I_{\mathrm{comp}}$ (= 1 mA). There were wide distributions in the RESET and SET voltages. The inset shows a typical $I$–$V$ curve near the RESET voltage of the NiO film. $R_o$ is the linear resistance value, obtained from the linear fit to the low $V$ region.

FIG. 2. (Color online) $R_o$ dependences of (a) RESET voltage, $V_R$, and (b) RESET current, $I_R$. At $R_{\mathrm{co}}$ (~30 Ω), the power law dependences for both $V_R$ and $I_R$ became varied. To clearly demonstrate how the power law varied with $R_o$, we selected eight data points. The selected $I$–$V$ curves are in the (c) $R_o < R_{\mathrm{co}}$ and (d) $R_o > R_{\mathrm{co}}$ regimes.

FIG. 3. (Color online) $R_o$ dependences of the third harmonic coefficient, $B_{3f}$. Note that $R_{\mathrm{co}}$ also separated the power law dependences for $B_{3f}$.

FIG. 4. (Color online) (a) A schematic of clusters connections in classical percolation

theory. $\xi$ is the coherence length, i.e., the average distance between nodes. Also shown are proposed schematic diagrams for the connectivity of conducting filaments in the regimes of (b) $R_o < R_{\mathrm{co}}$ (i.e., $\xi < L$, the sample size) and (c) $R_o > R_{\mathrm{co}}$ (i.e., $\xi > L$). The conducting filaments are multiply and singly connected in (b) and (c), respectively.

***Figure 1***

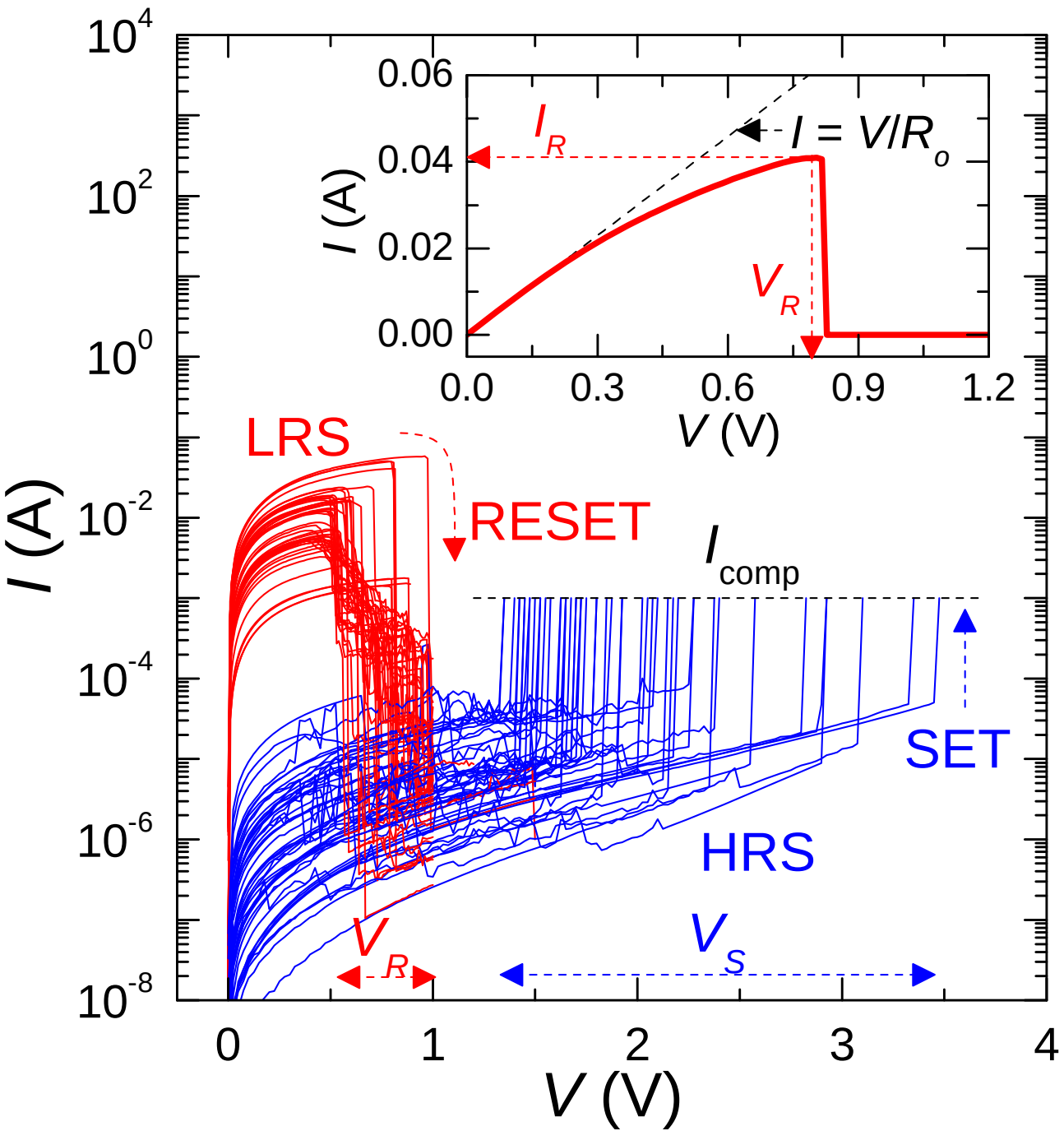

***Figure 2***

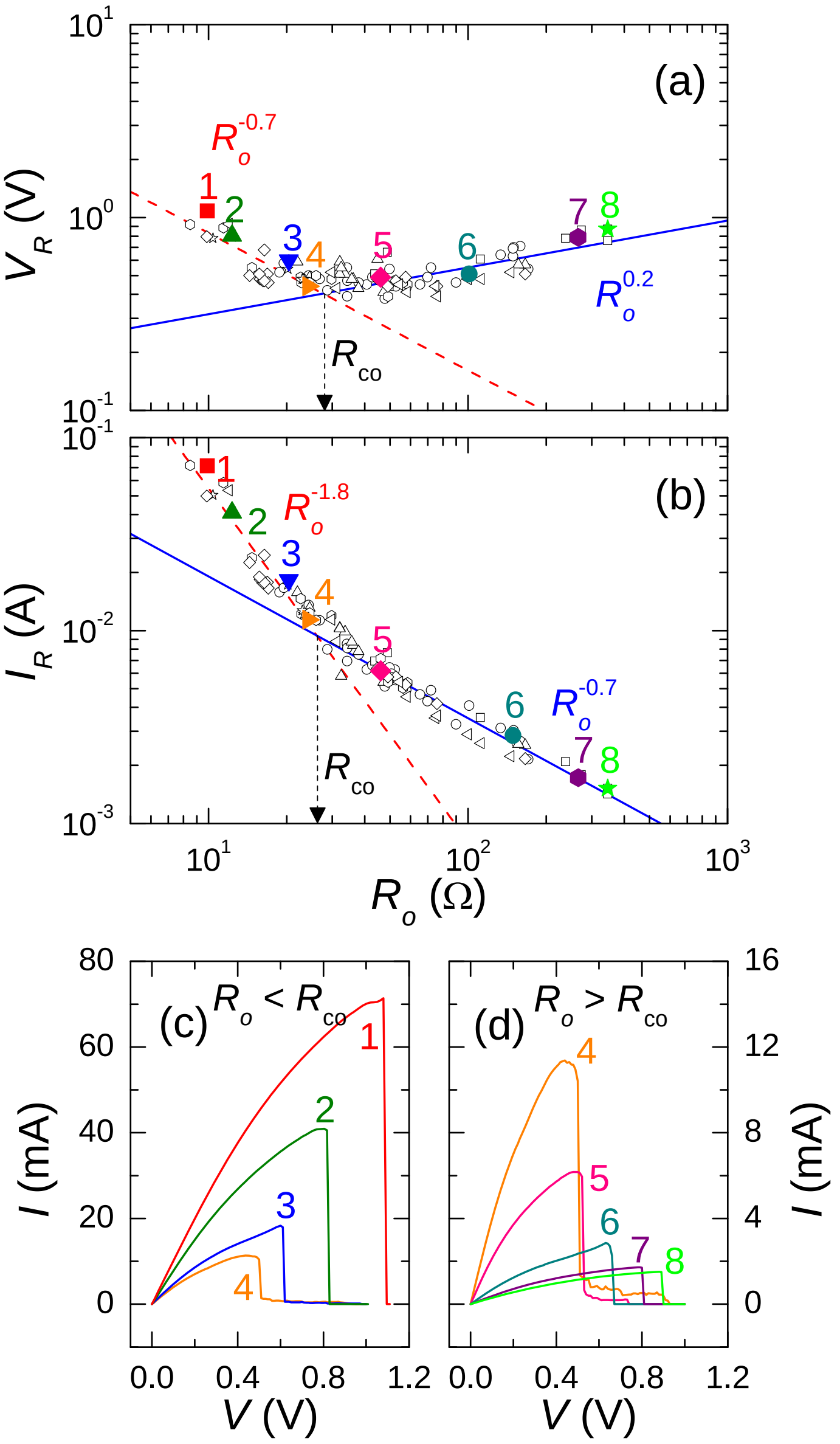

***Figure 3***

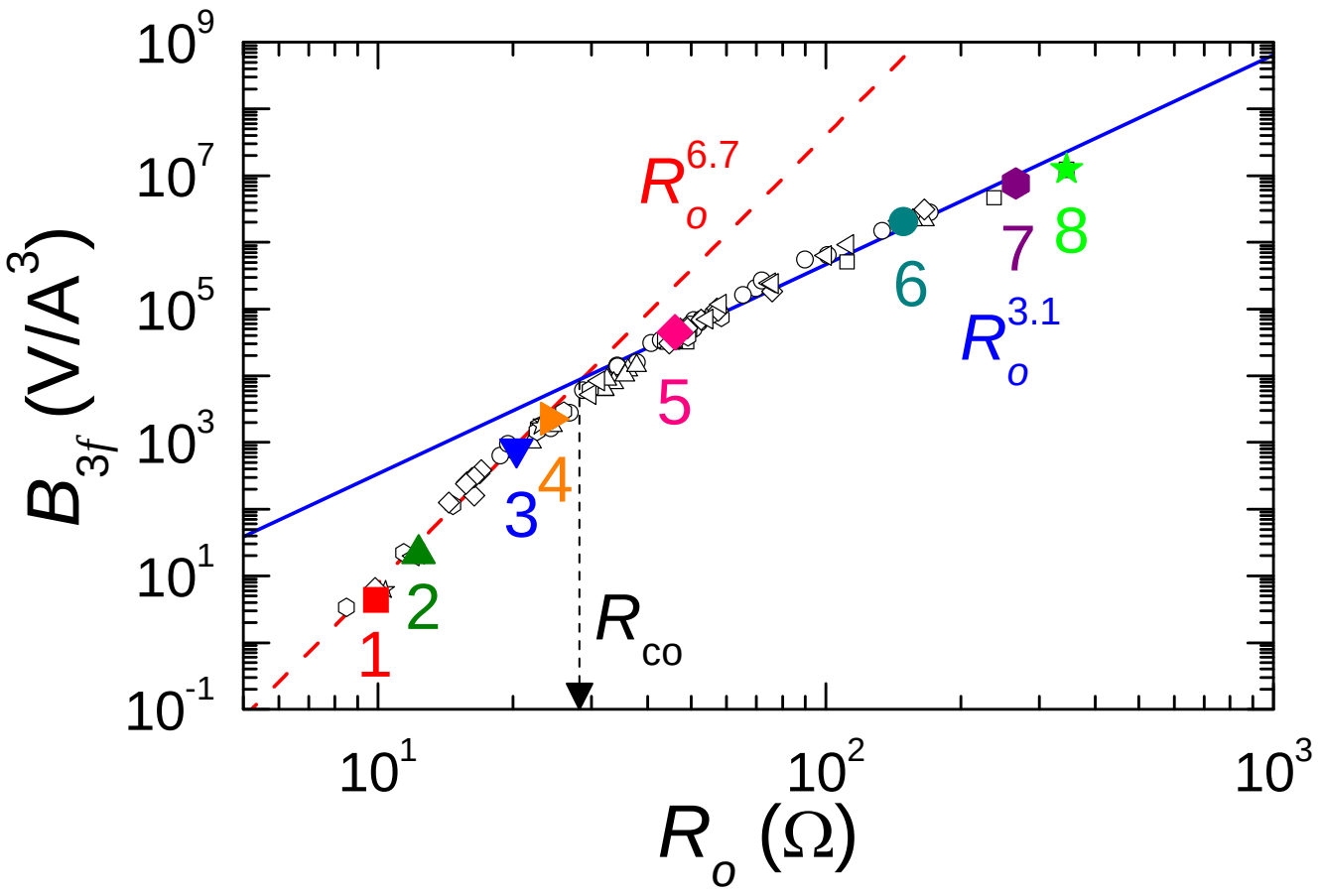

***Figure 4***

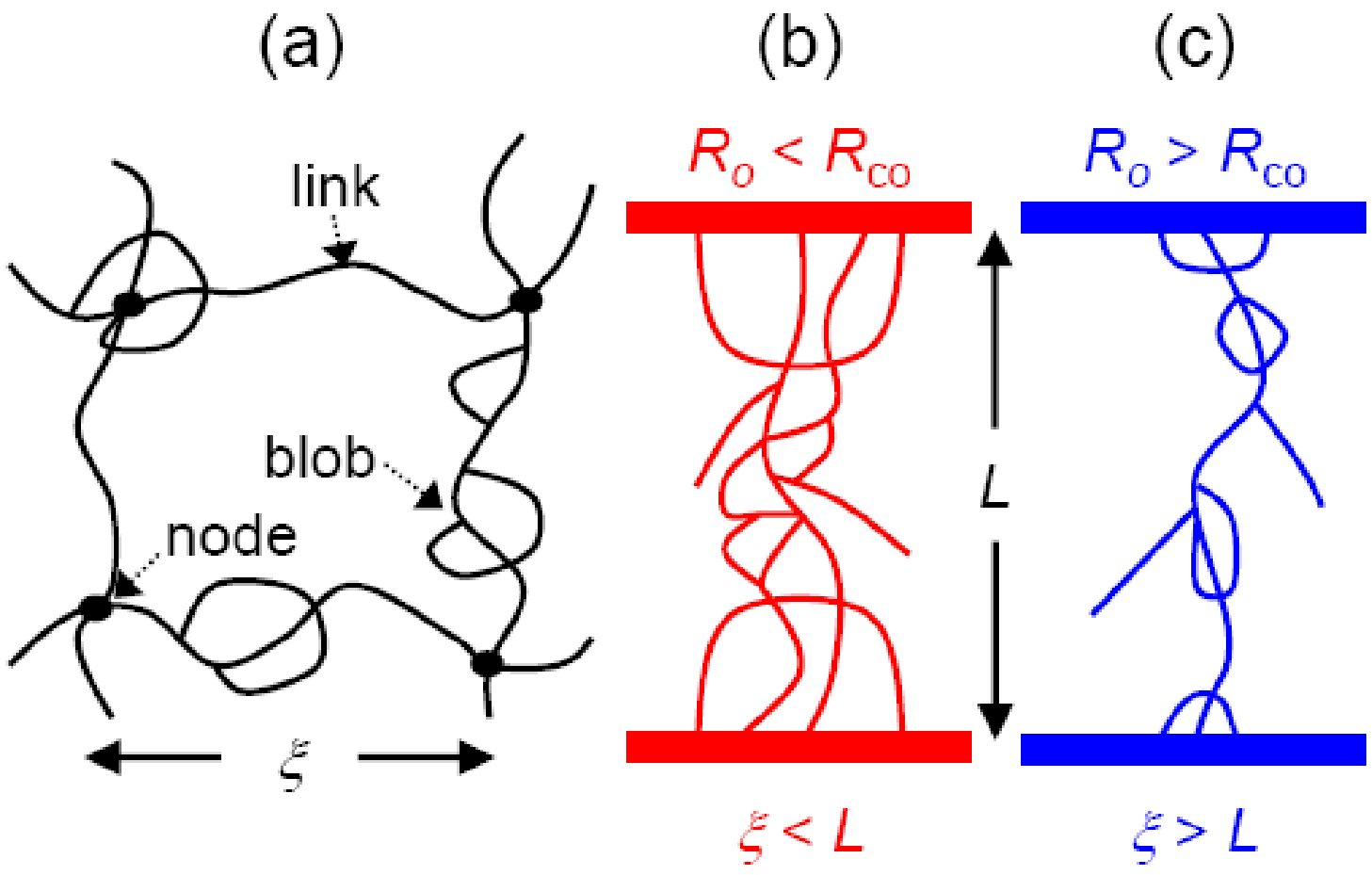